\documentclass{article}

\usepackage{axodraw}
\usepackage{graphicx}

\def\ba{\begin{eqnarray}}
\def\ea{\end{eqnarray}}
\def\tlscbf{\begin{picture}(4,4)(4,4)
\Photon(4,15)(26,15)2 4
\CArc(15,15)(10,0,180)
\CArc(15,15)(10,180,360)
\CArc(15,15)(12,0,180)
\CArc(15,15)(12,180,360)
\end{picture}}

\def\tlspbf{\begin{picture}(4,4)(4,4)
\Photon(4,15)(26,15)2 4
\CArc(15,15)(10,0,180)
\CArc(15,15)(10,180,360)
\CArc(15,15)(11,0,180)
\CArc(15,15)(11,180,360)
\CArc(15,15)(12,0,180)
\CArc(15,15)(12,180,360)
\end{picture}}

\def\tlscfr{\begin{picture}(4,4)(4,4)
\Photon(5,15)(25,15)2 4
\CArc(15,15)(10,0,180)
\CArc(15,15)(10,180,360)
\end{picture}}

\def\olscbf{\begin{picture}(4,4)(4,4)
\CArc(15,15)(10,0,180)
\CArc(15,15)(10,180,360)
\CArc(15,15)(12,0,180)
\CArc(15,15)(12,180,360)
\end{picture}}

\def\olscfr{\begin{picture}(4,4)(4,4)
\CArc(15,15)(10,0,180)
\CArc(15,15)(10,180,360)
\end{picture}}

\def\olscbfv{\begin{picture}(4,4)(4,4)
\CArc(15,15)(10,0,180)
\CArc(15,15)(10,180,360)
\CArc(15,15)(12,0,180)
\CArc(15,15)(12,180,360)
\Vertex(15,4)2
\end{picture}}

\def\olscfrv{\begin{picture}(4,4)(4,4)
\CArc(15,15)(10,0,180)
\CArc(15,15)(10,180,360)
\Vertex(15,5)2
\end{picture}}

\begin{document}

\title{Self-duality, Helicity and Higher-loop Euler-Heisenberg Effective Actions\footnote{Based on talks presented by G. Dunne at QTS3 (Cincinnati, OH, Sept 2003) and QFEXT03 (Norman, OK, Sept. 2003).}}

\author{Gerald V. Dunne$^1$ and Christian Schubert$^2$\\ \\
~$^1$Department of Physics\\
University of Connecticut\\
Storrs, CT 06269-3046 USA\\ \\
~$^2$Department of Physics\\
University of Texas Pan American\\
Edinburg, TX  78539 USA}

\date{}
\maketitle

\abstract{
The Euler-Heisenberg effective action in a self-dual background is remarkably simple at two-loop. This simplicity is due to the inter-relationship between self-duality, helicity and supersymmetry. Applications include two-loop helicity amplitudes, beta-functions and nonperturbative effects. The two-loop Euler-Heisenberg effective Lagrangian for QED in a self-dual background field is naturally expressed in terms of one-loop quantities. This mirrors similar behavior recently found in two-loop amplitudes in N=4 SUSY Yang-Mills theory.}

\section{One-Loop Euler-Heisenberg Effective Action}

In computing effective actions in gauge theories and in gravity, one can use the external field as a probe of the vacuum structure of the quantum theory\cite{dittrich}. A great deal has been learned from the seminal work of Heisenberg and Euler\cite{he}, who already in 1936 produced the paradigm for this entire approach by computing the nonperturbative, renormalized, one-loop effective action for QED in a background of constant field strength $F_{\mu\nu}$.
This special soluble case of a constant field strength leads to several important insights and applications:

$\bullet$ \underline{Light-Light scattering.} The Euler-Heisenberg effective action is nonlinear in the electromagnetic fields, and the quartic and higher terms represent new nonlinear interactions, the first of which is light-light scattering, which does not occur in the tree level Maxwell action. Expanding the Euler-Heisenberg answer to quartic order we find
\ba
S^{(1)}=\frac{e^4}{360 \pi^2 m^4}\int d^4x\,
\left[(\vec{E}^2-\vec{B}^2)^2+7(\vec{E}\cdot\vec{B})^2\right]+\dots
\ea
which gives the low energy limit (since the field strength was constant) of the amplitude for light-light scattering.

$\bullet$ \underline{Low-energy effective field theory.} In more modern language, the expansion of the Euler-Heisenberg expression in powers of the electromagnetic fields is an example of a low-energy effective action, which describes the interactions of the light  degrees of freedom which remain after integrating out the massive electron degrees of freedom. Indeed, the explicit expansion for a background magnetic field is 
\begin{eqnarray}
{\mathcal L}^{(1)}=-\hbar c
\frac{2}{\pi^2}\left(\frac{mc}{\hbar}\right)^4 \sum_{n=0}^\infty
\frac{2^{2n}{\mathcal B}_{2n+4}}{(2n+4)(2n+3)(2n+2)}
\left(\frac{eB\hbar}{m^2 c^3}\right)^{2n+4}
\label{1loopseries}
\end{eqnarray}
where ${\mathcal B}_{2n}$ are Bernoulli numbers. The dimensionless expansion parameter is the only gauge and Lorentz invariant combination of the background field, $B^2$, divided by compensating powers of the heavy mass scale $m$. Physically, this is the ratio of the cyclotron energy scale of the magnetic field to the electron rest energy scale, or the square of the ratio of the Compton wavelength to the magnetic length. The  expansion also shows that the perturbative series is divergent, as the Bernoulli numbers grow factorially fast in magnitude.
 
$\bullet$ \underline{Pair-production in an electric field.} The presence of a background electric field accelerates and splits the virtual vacuum dipole pairs, leading to $e^+e^-$ particle production. This instability of the vacuum was realized already by Heisenberg and Euler in 1936, and later formalized in the language of QED by Schwinger\cite{schwinger}.  The leading imaginary part of the  one-loop effective Lagrangian in a weak electric field is
\begin{eqnarray}
Im\, {\mathcal L}^{(1)} \sim\, \frac{e^2 E^2}{8\pi^3}
\exp\left[-\frac{m^2\pi }{eE}\right]
\label{imag}
\end{eqnarray}
This imaginary part gives the rate of vacuum non-persistence due to pair production. The rate is more appreciable when the $E$ field approaches a critical value $E_c\sim \frac{m^2 c^3}{e\hbar}\sim 10^{16}\,{\rm Vcm}^{-1}$, where the work done accelerating a virtual pair apart by a Compton wavelength is of the order of the rest mass energy for the pair.
Even though the condition of a constant electric field is rather unrealistic, the result (\ref{imag}) provides the starting point for more detailed analyses which incorporate time-dependent electric fields\cite{brezin}.

$\bullet$ \underline{Strong-field limit.} The strong-field limit also contains important physical information, as will be discussed in detail  below. For example, for a strong background magnetic field,
\ba
{\mathcal L}^{(1)}\sim 
\left(\frac{e^2}{12\pi^2}\right)\frac{B^2}{2}\ln\left(\frac{eB}{m^2}\right)
\label{large}
\ea
and the coefficient tells us the first coefficient of the QED $\beta$-function.

\section{Two-Loop Euler-Heisenberg Effective Action}

In this talk, I ask: what do we know about such effective actions beyond the one-loop level, and what can we learn from them?

In principle, the computation of the two-loop Euler-Heisenberg effective Lagrangian is completely straightforward, as we only need to compute a single vacuum diagram with an internal photon line and a single fermion (or scalar) loop, where these propagators are those in the presence of the background field. These propagators have been known in closed-form for a long time\cite{fock,schwinger}. For example, the scalar propagator in coordinate space is ($\eta$ is a gauge-dependent phase factor)
\begin{eqnarray}
\hskip -.5cm G_{\rm scalar}(x,x^\prime)&=& -i
\frac{e^{-i\eta}}{(4\pi)^2}\int_0^\infty
\frac{dT}{T^2}\,\exp\left[-im^2
T-L(T)+\frac{i}{4}z\beta(T) z\right]\nonumber\\\nonumber\\
z_\mu&\equiv& x_\mu-x_\mu^\prime\nonumber\\
\beta_{\mu\nu}&\equiv & \left[eF\, \coth(eFT)\right]_{\mu\nu}\nonumber\\
L&\equiv& \frac{1}{2}{\rm tr}\,\ln\left(\frac{\sinh(eFT)}{eFT}\right)
\label{fockprop}
\end{eqnarray}
Here $F$ is the constant field strength matrix. The new feature that makes two-loop nontrivial is the necessity of mass renormalization. In a {\it tour de force}, Ritus\cite{ritus} found explicit expressions for the renormalized two-loop effective Lagrangian for both spinor and scalar QED. Unfortunately, these expressions are extremely complicated two-parameter integrals (roughly speaking, there is one parameter integral left for each spinor/scalar propagator), and it is difficult to extract detailed physical information beyond their leading behavior. This is the case even if one restricts to the less general case of a constant magnetic field or a constant electric field background \cite{ritus,dittrich,dunsch}.

The main new development\cite{ds1,ds2,ds3} is that if the background field is chosen to be self-dual, so that
\ba
F_{\mu\nu}=\tilde{F}_{\mu\nu}
\label{selfdual}
\ea
then the two-loop Euler-Heisenberg effective Lagrangians take the remarkably simple closed forms :
\begin{eqnarray}
{\mathcal L}_{\rm spinor}^{(2)}
&=&
-\alpha^2 \,\frac{f^2}{2\pi^2}\left[
3\,\xi^2 (\kappa)
-\xi'(\kappa)\right]
\label{2lsp}
\end{eqnarray}
\begin{eqnarray}
{\mathcal L}_{\rm scalar}^{(2)}
&=&
\alpha^2 \,\frac{f^2}{4\pi^2}\left[
\frac{3}{2}\,\xi^2 (\kappa)
-\xi'(\kappa)\right]
\label{2lsc}
\end{eqnarray}
where $f$ is the field strength, $\frac{1}{4}F_{\mu\nu}F^{\mu\nu}=f^2$, and $\kappa$ is the natural  dimensionless parameter
\begin{eqnarray}
\kappa\equiv \frac{m^2}{2e f} 
\label{kappa}
\end{eqnarray}
The ubiquitous function $\xi(\kappa)$ is simply related to the Euler digamma function $\psi(\kappa)$:
\begin{eqnarray}
\xi(\kappa)\equiv -\kappa\left(\psi(\kappa)-\ln(\kappa)+\frac{1}{
2\kappa}\right) = -\frac{1}{2}\int_0^\infty ds \, e^{-2 \kappa s}\left(\frac{1}{\sinh^2 s}-\frac{1}{s^2}\right)
\label{xi}
\end{eqnarray}
The subtraction of the first two terms of the asymptotic expansion of $\psi(\kappa)$ correspond to renormalization subtractions.

The dramatic simplicity of the results (\ref{2lsp},\ref{2lsc}), compared to Ritus's general expressions, raises two questions:

$\bullet$ why are these expressions so simple?

$\bullet$ why are these expressions so similar?

\noindent The answer lies in the three-way relationship between self-duality, helicity and supersymmetry.

\underline{Self-dual fields have definite helicity:}
\ba 
\sigma_{\mu\nu}F_{\mu\nu}\left(\frac{{\bf 1}+\gamma_5}{2}\right)=0
\ea
It is well-known that amplitudes for external field lines with like helicities are particularly simple, in both gauge and gravitational theories, and these simplicities are thought to be deeply related to integrability conditions and the self-dual Yang-Mills equations\cite{mahlon,bardeen}. Since the effective action for a self-dual field is the generating function for like-helicity amplitudes, it is consistent that the effective action in a self-dual background should be simple. However, most of the amplitude results are for massless particles, while here we see a generalization to massive particles. For example, a new prediction is that for massive QED the low energy limit of the two-loop amplitudes with all external helicities $+$ behave as:
\begin{eqnarray}
\Gamma^{(2)}[k_1,\epsilon_1^+;k_2,\epsilon_2^+;
\dots;k_N,\epsilon_N^+]=\alpha\pi\frac{(2e)^N}{(4\pi)2 m^{2N-4}}\,
c^{(2)}_{N/2}\,\, \chi_N
\label{2lhel}
\end{eqnarray}
where the $\chi_N$ are standard helicity-basis factors\cite{ds3}, and the numerical coefficients $c_n^{(2)}$  come from the weak field expansion of the two-loop effective Lagrangians for a self-dual background:
\begin{eqnarray}
c^{(2)}_n =
\frac{1}{(2\pi)^2}\left\{
\frac{2n-3}{2n-2}\,{\mathcal B}_{2n-2}
+\frac{3}{2}\sum_{k=1}^{n-1}
\frac{{\mathcal B}_{2k}}{2k}
\frac{{\mathcal B}_{2n-2k}}{(2n-2k)} \right\} 
\nonumber
\end{eqnarray}
These results have since been generalized\cite{louise} to all helicity combinations (up to 10 external legs) by making an explicit helicity expansion of Ritus's general result.

\underline{Self-dual fields and SUSY:} self-duality implies that the corresponding Dirac operator has a
 quantum mechanical supersymmetry. Therefore, apart from zero modes, the Dirac operator has the same spectrum (but with a multiplicity of 4) as the corresponding scalar Klein-Gordon operator\cite{thooft,jackiw}. At one-loop this has the simple consequence  
\ba
{\mathcal L}^{(1)}_{\rm spinor}=-2{\mathcal L}^{(1)}_{\rm scalar}+\frac{1}{2} 
\left(\frac{ef}{2\pi}\right)^2\, \ln\left(\frac{m^2}{\mu^2}\right)
\ea
where $N_0=\left(\frac{ef}{2\pi}\right)^2$ is the zero mode number density.
At two-loop the effective action is not simply a log determinant, but the quantum mechanical SUSY of the Dirac operator relates the spinor propagator to the scalar propagator via simple helicity projections, which means that the two-loop effective action can be written as the sum of two terms involving matrix elements of the scalar propagator, with different numerical coefficients in the case of spinor and scalar QED. This explains why (\ref{2lsp}) and (\ref{2lsc}) have similar forms\cite{dgs}.

A more prosaic reason for the simplicity of the two-loop expressions (\ref{2lsp}) and (\ref{2lsc}) is that for a self-dual field, $F_{\mu\nu}F_{\nu\rho}=-f^2 \delta_{\mu\rho}$, which dramatically simplifies the propagators of spinors or scalars in the background field. For example, for a scalar particle, the Lorentz matrix structure in (\ref{fockprop}) disappears, leaving
\begin{eqnarray}
G_{\rm scalar}(x,x^\prime)=\left(\frac{ef}{4\pi}\right)^2\int_0^\infty
\frac{dt}{\sinh^2(e f t)}\,e^{ -m^2 t -\frac{ef}{4}(x-x^\prime)^2 
\coth(e f t)}
\label{xmprop}
\end{eqnarray}
\begin{eqnarray}
G_{\rm scalar}(p)=\int_0^\infty \frac{dt}{\cosh^2(e f t)}\,e^{ -m^2 t
-\frac{p^2}{ef} \tanh(e f t)}
\label{pmprop}
\end{eqnarray}
The simplification is even more dramatic for massless particles, where for scalars
\begin{eqnarray}
G_{\rm scalar}(x,x^\prime)=\frac{e^{
-\frac{ef}{4}(x-x^\prime)^2 }}{4\pi^2(x-x^\prime)^2}\qquad ; \qquad
G_{\rm scalar}(p)=\frac{1-e^{  -\frac{p^2}{ef} }}{p^2}
\label{zeromassprops}
\end{eqnarray}

\underline{QED $\beta$-functions.} Another application of background field propagators is to the computation of the $\beta$-function, which characterizes the scale dependence of the running coupling. This is a much simpler problem than finding the full renormalized effective Lagrangians (\ref{2lsp}, \ref{2lsc}), as we only need the leading strong field behavior of ${\mathcal L}$.
In this approach, one uses the external field as a probing scale, instead of the external momentum of a self-energy diagram (which is the usual text-book approach). 
The Gell-Mann Low renormalization group argument for $\Pi_{\mu\sigma}(q^2,\mu^2)$ can be adapted to the effective Lagrangian ${\mathcal L}(eF,\mu^2)$ to relate the strong-field limit of ${\mathcal L}$ to the $\beta$-function coefficients\cite{ritus}. A similar idea was used at two loop by Shifman and Vainshtein to compute the QED $\beta$-function using the operator product expansion\cite{shifman}, and has recently been extended to the three-loop QCD $\beta$-function\cite{bornsen}. There is an immediate combinatorial advantage to this background field approach, as there are many fewer diagrams at a given order, since there are no external lines. Furthermore, each diagram has fewer vertices and propagators. The price, of course, is that the spinor (or scalar) propagators are not free ones, but are in the background field. However, as we have seen in (\ref{xmprop},\ref{pmprop},\ref{zeromassprops}), in a self-dual background the propagators are still rather simple, so this potentially tips the balance back in favor of this approach.

At n-loop order, the $\beta$-function coefficient $\beta_n$ is related to the coefficient $b_n$ from the strong field limit
\begin{eqnarray}
{\mathcal L}^{(n)}\sim b_n \, F^2\, \ln \left(\frac{e|F|}{m^2}\right)
\label{strong}
\end{eqnarray}
For example, for a background magnetic field, from the Euler-Heisenberg and Ritus expressions we find the leading strong field behavior:
\begin{eqnarray}
{\mathcal L}^{(1)}_{\rm scalar}\sim \frac{e^2}{48\pi^2}\frac{B^2}{2}\ln\left(\frac{eB}{m^2}\right)
\qquad, \qquad
{\mathcal L}^{(1)}_{\rm spinor}&\sim& \frac{e^2}{12\pi^2}\frac{B^2}{2} \ln\left(\frac{eB}{m^2}\right)
\nonumber\\\nonumber\\
{\mathcal L}^{(2)}_{\rm scalar}\sim \frac{e^4}{64\pi^4}\frac{B^2}{2} \ln\left(\frac{eB}{m^2}\right)
\qquad , \qquad
{\mathcal L}^{(2)}_{\rm spinor}&\sim& \frac{e^4}{64\pi^4}\frac{B^2}{2} \ln\left(\frac{eB}{m^2}\right)
\label{strongb}
\end{eqnarray}
From the coefficients we can read off the QED $\beta$-functions :
\begin{eqnarray}
\beta_{\rm scalar}=\frac{e^3}{48\pi^2}+\frac{e^5}{64\pi^4}+\dots
\qquad ; \qquad
\beta_{\rm spinor}=\frac{e^3}{12\pi^2}+\frac{e^5}{64\pi^4}+\dots
\label{betas}
\end{eqnarray}
Since in some sense the self-dual background is the  "simplest", we now test this approach for a self-dual field. From the one-loop Euler-Heisenberg expression, and the closed-form two-loop expressions (\ref{2lsp}, \ref{2lsc}) we find
\begin{eqnarray}
{\mathcal L}^{(1)}_{\rm scalar}\sim 
\frac{e^2}{48\pi^2}f^2 \ln\left(\frac{ef}{m^2}\right) \qquad, \qquad
{\mathcal L}^{(1)}_{\rm spinor}&\sim& -\frac{e^2}{24\pi^2}f^2 \ln\left(\frac{ef}{m^2}\right)
\nonumber\\\nonumber\\
{\mathcal L}^{(2)}_{\rm scalar}\sim 
\frac{e^4}{64\pi^4}f^2 \ln\left(\frac{ef}{m^2}\right)
\qquad , \qquad
{\mathcal L}^{(2)}_{\rm spinor}&\sim& -\frac{e^4}{32\pi^4}f^2 \ln\left(\frac{ef}{m^2}\right)
\nonumber
\end{eqnarray}
So the argument works for scalar QED but fails for spinor QED, even at one-loop. The reason is that in a self-dual background spinor QED has zero modes, and these preclude the massless limit on which is based the argument\cite{ritus} relating the $\beta$-function and the strong field limit of ${\mathcal L}$. This argument must be modified to account for the zero modes\cite{dgs}.  At one loop the mismatch is due to the infrared divergence of the unrenormalized ${\mathcal L}_{\rm spinor}$, as is familiar from instanton physics\cite{thooft,jackiw}. But at two loop there is no IR divergence and the zero modes actually enter via the mass renormalization\cite{dgs}. Thus, to continue this strategy to $n^{th}$ order we find that for scalar QED one simply needs the coefficient of the leading single-log term as in (\ref{strong}), while for spinor QED one also needs the $(n-1)^{th}$ loop anomalous mass dimension coefficient.

\underline{Beyond two loops:} To go beyond two loops one should take advantage of the great progress that has been made recently in understanding the structure of higher-loop quantum field theory (without background fields). The general strategy is to manipulate diagrams to reduce the number to a set of so-called "master diagrams" which need to be computed. This has led, for example, to many new two-loop results for QCD scattering amplitudes\cite{glover}. Kreimer\cite{dirk} and collaborators have discovered an elegant Hopf algebra structure underlying the seemingly messy jumble of ordinary Feynman diagram perturbation theory. I conclude this talk with some brief comments and speculations about how these techniques might be extended to incorporate background fields. 

It is a simple exercise in free QED (i.e. no background field) to show that in 4 dimensions
\ba
{\rm spinor\,\, QED} : \quad \mbox{\tlscfr} \quad &=& -6\,e^2\,\left[\quad \mbox{\olscfr}\quad \right]^2 
\label{sp2}\\
{\rm scalar\,\, QED} : \quad\mbox{\tlscfr} \quad &=& \frac{3}{2}\,e^2\,\left[ \quad \mbox{\olscfr}\quad \right]^2
\label{sc2}
\ea
(The loop on the RHS is a scalar loop in each case.)  Thus, the two loop vacuum diagram can be expressed in terms of a simpler one loop diagram. These results can either be derived by computing each side using dimensional regularization, or a quicker proof follows from an integration-by-parts argument. Now reconsider the two-loop effective Lagrangian results (\ref{2lsp}) and (\ref{2lsc}) in this diagrammatic language. For a self-dual background, the scalar propagator loops are evaluated using (\ref{pmprop}), and turn out to be simply related to the $\xi(\kappa)$ function:
\ba
\mbox{\olscbf}\quad -\quad\mbox{\olscfr} \quad&\equiv &\int \frac{d^4p}{(2\pi)^4}\left[G(p)-G_0(p)\right]=-\frac{m^2}{(4\pi)^2} \,\frac{ \xi(\kappa)}{\kappa}
\label{xiint}\\
\mbox{\olscbfv}\quad -\quad\mbox{\olscfrv} \quad &\equiv &\int \frac{d^4p}{(2\pi)^4}\left[(G(p))^2-(G_0(p))^2\right]= \frac{\xi^\prime(\kappa)}{(4\pi)^2}
\label{xiprint}
\ea
Here the double lines refer to scalar propagators in the self-dual background and the single line is the free scalar propagator, while the dot on a propagator refers to the propagator being squared.

Thus, the results (\ref{2lsp}) and (\ref{2lsc}) for the renormalized two-loop effective Lagrangians can be written diagrammatically as
\ba
{\rm spinor:} \left[\,\,\,\,\,\,\mbox{\tlspbf}\,\,\,\, - \,\,\, \mbox{\tlscfr}\,\,\,\,\right]&=&-6\,e^2\, \left[\,\,\,\,\, \mbox{\olscbf}\,\,\,\, - \,\,\,\, \mbox{\olscfr}\,\,\,\, \right]^2 +\frac{(ef)^2}{2\pi^2} \left[\,\,\,\,\, \mbox{\olscbfv}\,\,\,\, -\,\,\,\, \mbox{\olscfrv}\,\,\,\,\right]
\label{2lspdig}\\
\nonumber\\
{\rm scalar:} \left[\,\,\,\,\, \mbox{\tlscbf}\,\,\,\,- \,\,\,\,\mbox{\tlscfr}\,\,\,\,\right]&=&\frac{3}{2}\,e^2\, \left[\,\,\,\,\, \mbox{\olscbf}\,\,\,\, - \,\,\,\, \mbox{\olscfr}\,\,\,\, \right]^2 -\frac{(ef)^2}{4\pi^2}  \left[\,\,\,\,\,  \mbox{\olscbfv}\,\,\,\, -\,\,\,\,  \mbox{\olscfrv}\,\,\,\, \right]
\label{2lscdig}
\ea
The comparison with the free vacuum diagram relations (\ref{sp2}, \ref{sc2}) is striking. 
The entire effect of the background field is quite minimal, and the two loop diagrams can still be expressed simply in terms of one-loop diagrams. Note that this is a highly nontrivial result, as both mass and charge renormalization have been taken into account in (\ref{2lspdig}, \ref{2lscdig}). If an algebraic, rather than a direct computational, understanding could be found for these relations, as has been done in the simpler theories without background fields, then the possibility of extending to higher loops becomes much greater.  A first step in this direction has been taken in Ref.\cite{sddr}, which translates the results of \cite{ds1,ds2,ds3,dgs} into dimensional regularization, the standard regularization technique for higher-loop analyses. Another tantalizing relation is the structural similarity between the results (\ref{2lspdig}, \ref{2lscdig}) for the effective Lagrangians in a self-dual background, and recent results relating two-loop and one-loop amplitudes in maximally SUSY Yang-Mills theory\cite{abdk}. This is a very different theory, and yet the two-loop amplitudes appear as squares of one-loop amplitudes, plus a one-loop correction, which is reminiscent of (\ref{2lspdig}, \ref{2lscdig}).

\noindent{\bf Acknowledgment:} GD thanks the Rockefeller Foundation for a Summer 2003 Residency at the Bellagio Center and the DOE for support through Grant  DE-FG02-92ER40716.


\begin{thebibliography}{99}

\bibitem{dittrich}
W.~Dittrich and M.~Reuter,
{\it Effective Lagrangians In Quantum Electrodynamics},
Springer Lect.\ Notes Phys.\  {\bf 220}  (1985);
W.~Dittrich and H.~Gies,
{\it Probing the quantum vacuum. Perturbative effective action approach in quantum electrodynamics and its applications},
Springer Tracts Mod.\ Phys.\  {\bf 166} (2000).

\bibitem{he}
W. Heisenberg and H. Euler, 
Z. Phys. {\bf 98} (1936) 714;
V. Weisskopf, 
\emph{Kong.\ Dans.\ Vid.\ Selsk.\ Math-fys.\ Medd.} {\bf XIV} No.~6 (1936).

\bibitem{schwinger}
J. Schwinger, 
``On gauge invariance and vacuum polarization'', 
Phys. Rev. {\bf 82} (1951) 664.

\bibitem{fock}
V.~Fock,
``Proper Time In Classical And Quantum Mechanics,''
Phys.\ Z.\ Sow. {\bf 12}, 404 (1937);
Y.~Nambu,
``The Use Of The Proper Time In Quantum Electrodynamics,''
Prog.\ Theor.\ Phys.\  {\bf 5}, 82 (1950).

\bibitem{brezin}
E.~Brezin and C.~Itzykson,
``Pair Production In Vacuum By An Alternating Field,''
Phys.\ Rev.\ D {\bf 2}, 1191 (1970);
V.~S.~Popov and M.~S.~Marinov,
``E+ E- Pair Production In Variable Electric Field,''
Yad.\ Fiz.\  {\bf 16}, 809 (1972);
``Electron - Positron Pair Creation From Vacuum Induced By Variable Electric Field,''
Fortsch.\ Phys.\  {\bf 25}, 373 (1977).

\bibitem{ritus}
V. I. Ritus, 
``Lagrangian of an intense electromagnetic field and quantum electrodynamics at short distances'', 
Sov. Phys. JETP {\bf 42} (1975) 774,
 ``Connection between strong-field quantum electrodynamics with short-distance quantum electrodynamics'',
{\it ibid} {\bf 46} (1977) 423.
 ``The Lagrangian Function of an Intense Electromagnetic
Field'', in {\it Proc. Lebedev Phys. Inst.} Vol. {\bf 168}, {\it Issues
in Intense-field Quantum Electrodynamics}, V. I. Ginzburg, ed., (Nova
Science Pub., NY 1987).
 
 \bibitem{dunsch}
G.~Dunne and C.~Schubert,
``Two-loop Euler-Heisenberg QED pair-production rate,''
Nucl.\ Phys.\ B {\bf 564}, 591 (2000)
[arXiv:hep-th/9907190].

\bibitem{shifman}
M.~A.~Shifman and A.~I.~Vainshtein,
``Operator Product Expansion And Calculation Of The Two Loop
Gell-Mann-Low Function,'' 
Sov.\ J.\ Nucl.\ Phys.\  {\bf 44}, 321 (1986).

\bibitem{bornsen}
J.~P.~Bornsen and A.~E.~van de Ven,
``Three-loop Yang-Mills beta-function via the covariant background field  method,''
Nucl.\ Phys.\ B {\bf 657}, 257 (2003).
[arXiv:hep-th/0211246].

\bibitem{ds1}
G.~V.~Dunne and C.~Schubert,
``Closed-form two-loop Euler-Heisenberg Lagrangian in a self-dual background,'' 
Phys.\ Lett.\ B {\bf 526}, 55 (2002)
[arXiv:hep-th/0111134].

\bibitem{ds2}
G.~V.~Dunne and C.~Schubert,
``Two-loop self-dual Euler-Heisenberg Lagrangians. I: Real part and helicity amplitudes,'' 
JHEP {\bf 0208}, 053 (2002)
[arXiv:hep-th/0205004].

\bibitem{ds3}
G.~V.~Dunne and C.~Schubert,
``Two-loop self-dual Euler-Heisenberg Lagrangians. II: Imaginary part  and Borel analysis,'' 
JHEP {\bf 0206}, 042 (2002)
[arXiv:hep-th/0205005].

\bibitem{mahlon} 
M.~L.~Mangano and S.~J.~Parke,
``Multiparton Amplitudes In Gauge Theories,''
Phys.\ Rept.\  {\bf 200}, 301 (1991);
G. Mahlon, 
``One loop multiphoton helicity amplitudes'', 
Phys. Rev. D {\bf 49} (1994) 2197;
Z.~Bern, L.~J.~Dixon, M.~Perelstein and J.~S.~Rozowsky,
``Multi-leg one-loop gravity amplitudes from gauge theory,''
Nucl.\ Phys.\ B {\bf 546}, 423 (1999);
[arXiv:hep-th/9811140];
Z.~Bern, A.~De Freitas and L.~Dixon,
``Two-loop helicity amplitudes for gluon gluon scattering in QCD and  supersymmetric Yang-Mills theory,''
JHEP {\bf 0203}, 018 (2002);
[arXiv:hep-ph/0201161];
T.~Binoth, E.~W.~N.~Glover, P.~Marquard and J.~J.~van der Bij,
``Two-loop corrections to light-by-light scattering in supersymmetric QED,''
JHEP {\bf 0205}, 060 (2002).
[arXiv:hep-ph/0202266].

\bibitem{bardeen} 
W. Bardeen, 
``Self-dual Yang-Mills theory, integrability and multiparton amplitudes'', 
Prog. Theor. Phys. Suppl. {\bf 123} (1996) 1; 
D. Cangemi, 
``Self-duality and maximally helicity violating QCD amplitudes'', 
Int. J. Mod. Phys. {\bf A12}
(1997) 1215.
[arXiv:hep-th/9610021].

\bibitem{louise}
L.~C.~Martin, C.~Schubert and V.~M.~Villanueva Sandoval,
``On the low-energy limit of the QED N - photon amplitudes,''
Nucl.\ Phys.\ B {\bf 668}, 335 (2003).
[arXiv:hep-th/0301022].

\bibitem{thooft}
G.~'t Hooft,
``Computation Of The Quantum Effects Due To A Four-Dimensional Pseudoparticle,'' 
Phys.\ Rev.\ D {\bf 14}, 3432 (1976).
[Erratum-ibid.\ D {\bf 18}, 2199 (1978)].

\bibitem{jackiw}
R.~Jackiw and C.~Rebbi,
``Degrees Of Freedom In Pseudoparticle Systems,''
Phys.\ Lett.\ B {\bf 67}, 189 (1977);
``Spinor Analysis Of Yang-Mills Theory,''
Phys.\ Rev.\ D {\bf 16}, 1052 (1977).

\bibitem{dgs}
G.~V.~Dunne, H.~Gies and C.~Schubert,
``Zero modes, beta functions and IR/UV interplay in higher-loop QED,''
JHEP {\bf 0211}, 032 (2002)
[arXiv:hep-th/0210240].

\bibitem{glover}
E.~W.~N.~Glover,
``Progress in NNLO calculations for scattering processes,''
Nucl.\ Phys.\ Proc.\ Suppl.\  {\bf 116}, 3 (2003);
Z.~Bern,
``Recent progress in perturbative quantum field theory,''
Nucl.\ Phys.\ Proc.\ Suppl.\  {\bf 117}, 260 (2003).
[arXiv:hep-ph/0212406].

\bibitem{dirk}
D.~Kreimer,
``On the Hopf algebra structure of perturbative quantum field theories,''
Adv.\ Theor.\ Math.\ Phys.\  {\bf 2}, 303 (1998);
[arXiv:q-alg/9707029];
A.~Connes and D.~Kreimer,
``Renormalization in quantum field theory and the Riemann-Hilbert  problem. I: The Hopf algebra structure of 
 graphs and the main theorem,''
Commun.\ Math.\ Phys.\  {\bf 210}, 249 (2000);
[arXiv:hep-th/9912092];
D. Kreimer, {\it Knots and Feynman Diagrams}, (Cambridge Univ Press, 2000).

\bibitem{sddr} G.~ Dunne, ``Two-Loop Diagrammatics in a Self-Dual Background", hep-th/0311167.

\bibitem{abdk}
C.~Anastasiou, Z.~Bern, L.~Dixon and D.~A.~Kosower,
``Planar amplitudes in maximally supersymmetric Yang-Mills theory,''
arXiv:hep-th/0309040.

\end{thebibliography}
\end{document}